\documentclass[11pt,twoside]{article}
\usepackage{asp2004}
\usepackage{psfig}
\usepackage{epsf}
\usepackage{graphics}
\usepackage{natbib}
\usepackage{lscape}
\def\edcomment#1{\iffalse\marginpar{\raggedright\sl#1\/}\else\relax\fi}
\reversemarginpar

\begin{document}
\title{Cosmokinetics}
\author{R. D. Blandford, M. Amin, E. A. Baltz, K. Mandel \& P. J. Marshall}
\affil{KIPAC, Stanford University, Stanford, CA 94309}

\begin{abstract}
Our fundamental lack of understanding of the acceleration of the
Universe suggests that we consider a kinematic description. The
simplest formalism involves the third derivative of the scale
factor through a jerk parameter. A new approach is presented for
describing the results of astronomical observations in terms of
the contemporary jerk parameter and this is related to the
equation of state approach. Simple perturbative expansions about
$\Lambda$CDM are given.
\end{abstract}

\thispagestyle{plain}
\section{Approaches to Dark Energy}
The Universe is spatially flat, accelerating and has a subcritical
matter density. The evidence for this comes from:
\begin{itemize}
\item observation of acoustic peaks in the microwave background
radiation spectrum which effectively measures the angular diameter
distance $d_A$ to the last scattering surface \cite{Spergel2003}.
\item magnitudes of Type Ia
supernovae which provide the luminosity distance $d_L$ as a
function of redshift \cite{Riess2001}.
\item X-ray observations of rich clusters of
galaxies which measure the mean density of dark matter and also, with
less confidence, the distance--redshift relation \cite{Allen2004}.
\end{itemize}

Although the possibility that the Universe contains a repulsive
entity counteracting the attractive effects of gravity was
entertained by Einstein in the first, relativistic  cosmology, the
discovery of  acceleration was mostly a surprise. There have been
several approaches to rationalizing this discovery from a
contemporary perspective. The simplest, and probably still the
favorite description, is that the field equation is supplemented
by a term proportional to the metric tensor, which guarantees
covariance. The coefficient of proportionality (the cosmological
constant) is the single free parameter in the theory.

If we regard this term as an augmentation of the stress energy
tensor of cold matter and adopt a fluid description, then the
enthalpy must vanish and the pressure is the negative of the
energy density (roughly 0.7 nJ~m$^{-3}$). Alternatively, we can
regard this term as a necessary component of the geometrical
response of a spacetime manifold to a material source that only
becomes detectable on cosmological scales. Either choice leads to
the flat ``$\Lambda$CDM'' cosmology in which the cosmic time
varies with the scale factor $a=1/(1+z)$ according to the solution
of Lema\^{i}tre (1927) equation
\begin{equation}
\label{eq:scale}
H_0t(a)={2\over3(1-\Omega_m)^{1/2}}{\sinh^{-1}\left(a^{3}{(1-\Omega_m)}\over\Omega_m\right)^{1/2}}
\end{equation}
\cite{Bondi1952}, where $\Omega_m$ is the density parameter of
matter ($\rho=\Omega\rho_c$ where the critical density
$\rho_c=3H_0^2/8\pi G$) and $a$ is normalized to unity at the
present time\footnote{Eddington (1935) strongly advocated a
$\Lambda$CDM model on the grounds that it would prolong the life
of the Universe and that the existence of a cosmological constant
was necessary to explain the disparity between the atomic and
cosmological scales.}. Following Spergel {\it et al.} (2003), we
adopt $H_0=72$~km s$^{-1}$ Mpc$^{-1}$.

In order to test this description, it is necessary to introduce a
framework which embraces a larger class of models labeled by
measurable parameters. The overwhelmingly favourite choice of
model for ``dark energy'' is to suppose that the acceleration is
caused by a fluid which expands according to a modified adiabatic
law, so that  $PV^{1+w}$ is equal to a (negative) constant
\cite{{Steinhardt1997},{Turner1997}}. (This is reminiscent of a
polytropic description of a star with the important distinction
that it is the same material that undergoes the change as the
Universe expands as opposed to matter with different thermal
histories as is the case with a star.) The parameter $w$ ($=-1$
under $\Lambda$CDM) can be allowed to vary but we shall treat it
as constant for the moment. Given this prescription $a$ satisfies
a Friedmann differential equation (neglecting curvature)
\begin{equation}
\label{eq:friedmann} \dot a^2\equiv -2H_0^2V(a)=H_0^2\left[\Omega_m
a^{-1}+(1-\Omega_m)a^{-(1+3w)}\right]
\end{equation}

This approach can be confusing because, if we were to treat $w$ as
a pure number then small scale perturbations to the dark energy
fluid would grow exponentially with time for $w<0$. This is
clearly untenable. Instead $w$ is usually associated with a scalar
field theory with a more complex rule for handling perturbations.
Under these circumstances, it would seem preferable to
parameterize the field theory itself which will lead to a
different functional departure of the models from $\Lambda$CDM. An
analogy is that one invariably solves problems in electromagnetic
theory working from Maxwell's field equations rather than his
stress tensor. In practice small scale dark energy perturbations
propagate with the speed of light in most, though not all,
models~\cite{Liu2004}.

A huge variety of scalar field models have been discussed
(e.g.Caldwell {\it et.al.} 1998; Peebles \& Ratra 1988). However,
there is no guarantee that the acceleration of the Universe has
anything to do with a scalar field. Other explanations include the
notion that the Universe has extra dimensions \cite{Dvali1998}.
The scalar field is probably the best bet although it should be
remembered that no such fundamental fields have been detected
experimentally as yet. Electromagnetic and gravitational fields
were thought to be scalar, but turned out not to be. To make this
point, consider a blind astrophysicist practising her craft during
the radiation era. She might have measured that the average
spacing of protons increased as the square root of time and have
inferred that the expansion was driven by a scalar field with an
exponential potential. She would have been hopelessly wrong.
\section{Kinematic Approach}
\subsection{Jerk Parameter}
In view of the huge diversity of dynamical models on offer, there
may be some merit in the observational cosmologist taking an
unprejudiced kinematical approach. This has been historically
productive. Galileo, who originally argued theoretically that the
speed of a falling body increased in proportion to distance,
determined the true law by empirical, kinematic measurement
\cite{Dugas1955}. Modern cosmology began with the discovery of the
expansion of the universe, quantified by the Hubble parameter
$H(t)=\dot a/a$.  In more recent times the pursuit of
observational cosmology was dominated by the measurement of the
contemporary deceleration parameter $q_0$ where $q(t)=-\ddot
aa/\dot a^2$ \cite{Sandage1975}. In terms of the scale factor and
Hubble parameter $q=-d\,\ln aH/d\,\ln a$.

As the Universe was once decelerating and is now accelerating, it
is useful to consider the third derivative of~$a$. A convenient
quantity to use is the dimensionless jerk \cite{Blandford2004}
\begin{equation}
\label{eq:jerk} j={\stackrel{\ldots}{a}a^2\over\dot a^3}
\end{equation}
with its current value denoted by $j_0$. One reason why this is
convenient is that $j=1$ for $\Lambda$CDM, the baseline model
about which we are perturbing.  In terms of the Hubble parameter
$j=(a^2H^2)''/2H^2$.  Different formalisms involving the third
derivative have been discussed independently by
\cite{{Sahni2003},{Visser2004}}.

\subsection{Dynamical considerations}
Although, we eschew dynamics, $j$ can be interpreted in fluid
terms, through the easily derived relation $j=1-4\pi\dot
P/H^3=1+(a^4\rho')'/2\rho a^2$, where $P$ is the pressure, $\rho$
the total energy density while we ignore spatial curvature
(analagously $-q=1+a\rho'/2\rho$).  The expressions involving $H$
rather than $\rho$ above are valid even considering curvature. For
a slow rolling scalar field  $\phi$ and potential $U(\phi)$ we get
$j\approx1-4\pi U'(\phi)^2 /3H^4$.

\subsection{Hubble Parameter}
We solve Eq.~\ref{eq:jerk} by rewriting it in the form
\begin{equation}
\label{eq:kinetic} V''={2jV\over a^2}
\end{equation}
where $V(a)=-{\dot{a}^2}/2H_0^2=-{H^2(a)}{a^2}/2H_0^2$ and prime
denotes differentiation with respect to $a$. This equation has
power law solutions for constant $j$, but the only one that is of
interest is the $\Lambda$CDM solution with $j=1$. This is because
we seek solutions that approximate the Einstein-De Sitter case for
early times. We therefore take for our simplest, one parameter
extension of $\Lambda$CDM, the variation
\begin{equation}
\label{eq:djerk}
j(a)=1+j'a=1+(j_0-1)a
\end{equation}
In general, if we specify an expression for $j(a)$ and require
that the solution match an Einstein-De Sitter solution at early
time, then we can solve for the scale factor, $a(t)$, the Hubble
parameter, $H(t)$, and the deceleration parameter, $q(t)$ for all
time. (Specifically all we need to know is $j(a)$ and
$\lim_{a\rightarrow0}(a\dot a^2/H_0^2)$.) As is the case with the
$w$ formalism, it is possible to introduce extra parameters
through a Taylor expansion $j(a)=1+j'a+j''a^2/2+\dots$  However,
we shall not explore this generalization here, and instead
restrict ourselves to a jerk that is linear in the scale factor.
\begin{figure}
\begin{center}
\scalebox{0.7} {\includegraphics{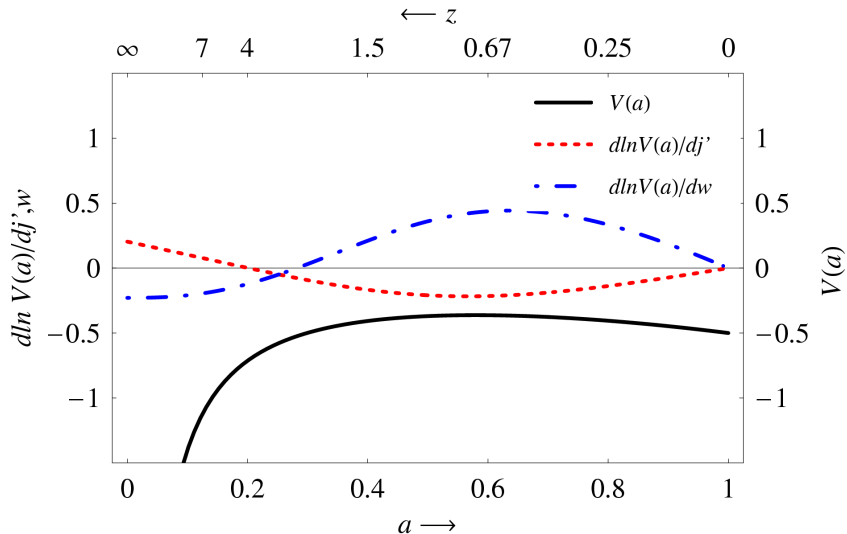}}
\caption{The effective potential $V(a)$, and the derivative $d\ln
V(a)/dj',dw$, evaluated at $j'=0$ {$(w=-1)$}.}
\label{fig:kin}
\end{center}
\end{figure}
Eq.~\ref{eq:kinetic} has the solution
\begin{equation}
\label{eq:jerksol} V(a) = -{1\over2}\left[ c_1 f_1 (8aj')/a + c_2
f_2 (8aj') a^2 \right]
\end{equation}
where
\begin{eqnarray}
\label{eq:bessel}
f_1(x)&=&x^{3/2}K_3(x^{1/2})/8=1-x/8+x^2/64+\dots\cr
f_2(x)&=&48x^{-3/2}I_3(x^{1/2})=1+x/16+x^2/640+\dots
\end{eqnarray}
and $c_{1,2}$ are constants.
\begin{figure}
\begin{center}
\scalebox{0.7}{\includegraphics{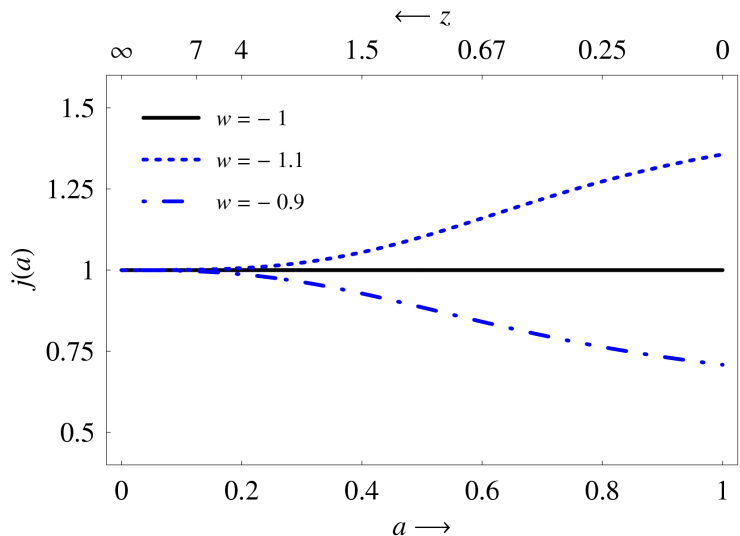}}
\caption{The kinematic jerk parameter $j(a)$ shown for three different
values  of the dark energy equation of state parameter $w$.}
\label{fig:jerk}
\end{center}
\end{figure}

\subsection{\boldmath${j}$ vs. \boldmath${w}$}
In order to compare the $j$ and $w$ prescriptions, we must compare
world models computed adopting similar boundary conditions. We
choose to fix the co-moving distance
${H_0}d(a_r)=H_0\int_{a_r}^1dt /a$ to the recombination
surface($a_r \approx 0$). Note that the distance to the
recombination surface, ${H_0}d(0)$ is in practice measured quite
accurately by the microwave background fluctuation spectra. In
everything that follows, we shall assume ${H_0}d(0)=3.4.$ This
condition allows us to solve for $\Omega_m$ as a function of $j'$
(or $w$). If we linearize in $j'$ and assume that matter evolves
in the usual way then in the $j$ model, $c_1=\Omega_m$ and $c_2 $
can be chosen such that $\ V(1)=-1/2$. The linearized expressions
are shown below for illustrative purposes.
\begin{eqnarray}
\label{eq:linearize}
V(a)&=&-{1\over2}\left[\Omega_m/a+(1-\Omega_m)a^2+j'\{(1-\Omega_m)a^3/2-(1-3\Omega_m)a^2/2-\Omega_m\}\right]\cr
V(a)&=&-{1\over2}\left[\Omega_m/a+(1-\Omega_m)a^2-3(w+1)(1-\Omega_m)a^2\ln
a\right]
\end{eqnarray}
In Fig.~\ref{fig:kin} we plot the exact solution for
$V(a)$ along with the derivative $d\ln V(a)/dj',dw$
evaluated at $j'=0$ ($w=-1$) to show the difference between the
$w$ and $j$ formalisms.

In general $d\ln X/dj'$ characterizes the deviation of an
observable $X$ from the ${\Lambda}$CDM model. This can be seen by
considering the following expansion:
\begin{equation}
\label{eq:fitting}
X(a,j') \approx X(a,0)\left(1+j'\frac{d\ln X}{dj'}{\bigg\vert}_{j'=0}\right)
\end{equation}
Note that this provides a
simple route to the generation of fitting formulae.
Substituting $w$ for $j'$ and expanding about the ${\Lambda}CDM$ value
of $w=-1$ gives the corresponding expression in the dynamical
formalism.

We can also compute $j(a)$ for a given $w$ model.
\begin{equation}
\label{eq:wtoj}
j(a)=1+{9w(1+w)(1-\Omega_m)\over2\{1-\Omega_m(1-a^{3w})\}}
\end{equation}
This relationship is plotted in Fig.~\ref{fig:jerk}.

\subsection{Distance}
\begin{figure}
\begin{center}
\scalebox{0.7}{\includegraphics{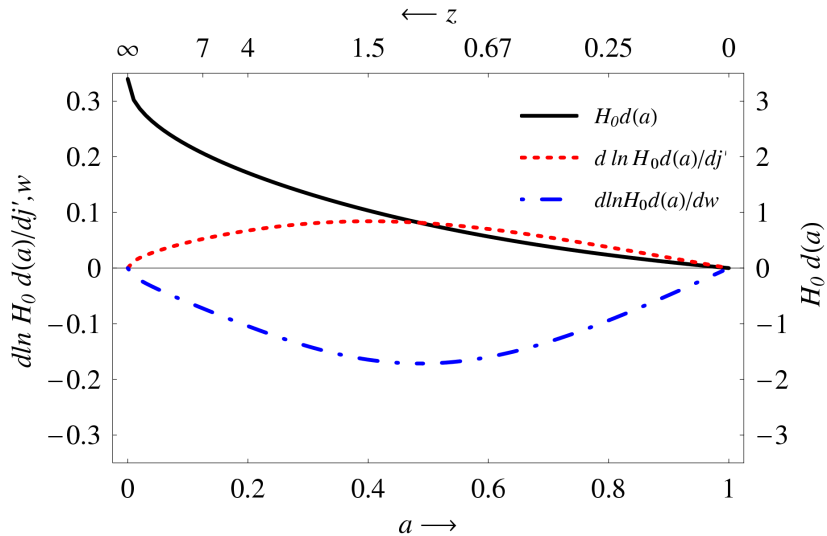}}
\caption{Co-moving distance $H_0d(a)$ and the derivative $d\ln
H_0d(a)/dj',dw$ evaluated at $j'=0$ ($w=-1$).}
\label{fig:dist}
\end{center}
\end{figure}

Many techniques for exploring the expansion of the Universe
utilize distance measurement through the angular diameter
$d_A\equiv ad$ or luminosity distance $d_L\equiv a^{-1}d$ or the
comoving volume $\mathcal{V}=4\pi d^3/3$. For $\Lambda$CDM,
\begin{equation}
\label{eq:eta}
H_0d(a)=H_0d(0)-{B[1/6,1/3;(1-\Omega_m)a^3/({\Omega_m+(1-\Omega_m
)a^3})]\over{3\Omega_m^{1/3}(1-\Omega_m)^{1/6}}}
\end{equation}
where $B[a,b;x]=\int_0^xdt\,t^{a-1}(1-t)^{b-1}$ is the incomplete
Beta function. The distances can be computed in the perturbed
models with the the boundary condition ${H_0}d(0)=3.4$. We
contrast the $j$ and $w$ formalisms in Fig.~\ref{fig:dist} where
we plot ${H_0}d$ and $d\ln {H_0}d/dj',dw$  evaluated at $j'=0(w=-1$) as
functions of $a$. The fractional changes in the co-moving volume
are three times these functions.
A simple fitting formula for ${H_0}d$ is given by combining Eq.\
\ref{eq:fitting} with
$d\ln {H_0}d/dj' \approx \sum_{n=1}^{4} c_n a^n$, where the coefficients
$c_n=\{0.54,-1.1,0.81,-0.26\}$.

\subsection{Curvature}
\begin{figure}
\begin{center}
\scalebox{0.7}{\includegraphics{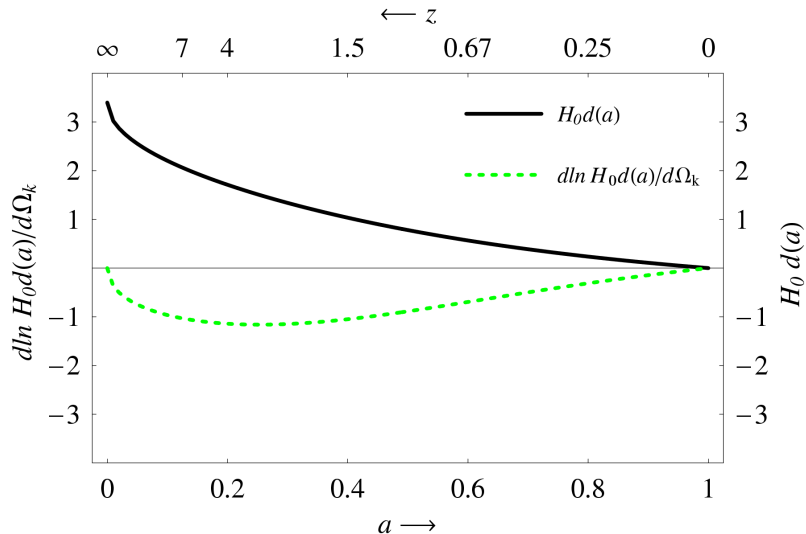}}
\caption{Co-moving distance $H_0d(a)$, and the derivative $d\ln
H_0d(a)/d\Omega_k$, evaluated at $\Omega_k=0, j'=0 (w=-1)$.}
\label{fig:curv}
\end{center}
\end{figure}

The WMAP observations have demonstrated that the Universe is quite
flat; any curvature can surely be treated as a perturbation and
will be very hard to detect in the face of the above effects.  We
therefore contrast the deviation expected from pure curvature in
the distance relation. Curvature with radius $R$ appears in
standard Friedmann cosmology in two related places. The first is
in the relation between the comoving radial coordinate $r=\int
dt/a$ and the distance $d=r+\Omega_kr^3/6+\dots$, and
$|\Omega_k|=1/R^2$, where the sign indicates positive or negative
spatial curvature ($\Omega_k$ negative is positive curvature).
Note that this effect is purely geometric, making no reference to
dynamics.  The second is in the dynamical equation,
Eq.~\ref{eq:friedmann}, which changes so that
$V=-\left[\Omega_ma^{-1}+(1-\Omega_m-\Omega_k)a^{-(1+3w)}+\Omega_k\right]/2$.
The jerk parameter can be derived from $j=1+(a^4\rho')'/2a^2\rho$
by defining $\rho_k=\Omega_k\rho_c a^{-2}$ and taking
$\rho\rightarrow\rho+\rho_k$.  For the pure $w$ model,
$j_0=1-\Omega_k+9(1-\Omega_m-\Omega_k)w(1+w)/2$. In
Fig.~\ref{fig:curv} we show $d\ln H_0d/d\Omega_k$ evaluated at
$\Omega_k=0,j'=0(w=-1)$.
\subsection{Time}
\begin{figure}\
\begin{center}
\scalebox{0.7}{\includegraphics{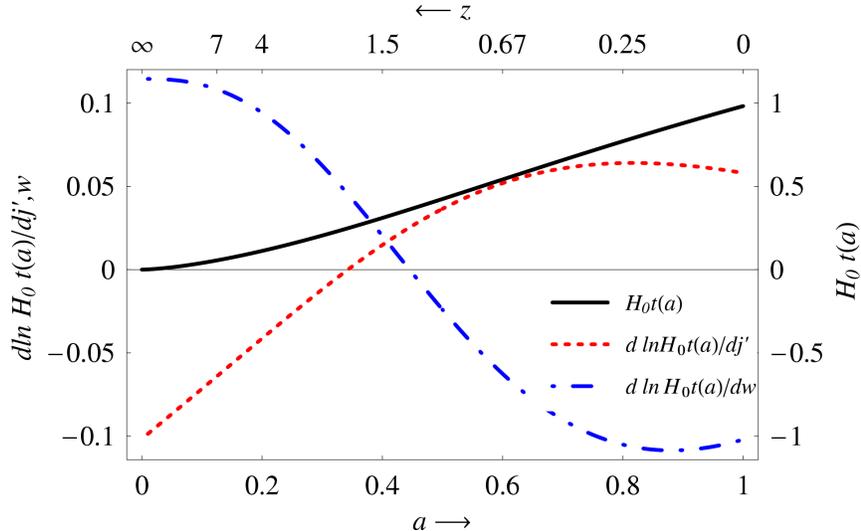}}
\caption{Cosmic time $H_0t(a)$, and the derivative $d\ln
H_0t(a)/dj',dw$, evaluated at $j'=0 (w=-1)$.}
\label{fig:time}
\end{center}
\end{figure}

At present we do not have very accurate age measurements but it is
still useful to compute the relative changes in the chronology
induced by our models. We obtain the time under $\Lambda$CDM from
Eq.~\ref{eq:scale}. The times for the $w,j$ modifications are
obtained by integrating Eq.~\ref{eq:linearize} subject to a third
boundary condition that $t(0)=0$. $d\ln H_0t/dj',dw$ for
$j'=0(w=-1)$ are plotted in Fig.~\ref{fig:time}.
\subsection{Growth of Perturbations}
The rate at which the perturbations that are observed directly in
the microwave background fluctuation spectra grow under the action
of Newtonian gravitational forces is quite sensitive to the
expansion rate of the Universe. If we ignore radiation, pressure
and the distinction between cold and baryonic matter, then a
relative matter density perturbation on a fixed co-moving scale,
denoted by  $\delta$, grows according to a second order
differential equation which can be cast in Sturm-Liouville form
\begin{equation}
\label{eq:growth}
 (a^2V^{1/2}\delta')'+{3\Omega_m\delta\over4aV^{1/2}}=0
 \end{equation}
We solve this differential  equation evolving the linear growth
transfer function for the $w$ and $j$ models from the epoch of
recombination  to the present day assuming $\delta(a_{r})=1$ and
$\delta'(a_{r})=1/a_{r}$. The difference between the $w$ and $j$
models is shown using $d\ln\delta/dj',dw$.
\begin{figure}\
\begin{center}
\scalebox{0.7}{\includegraphics{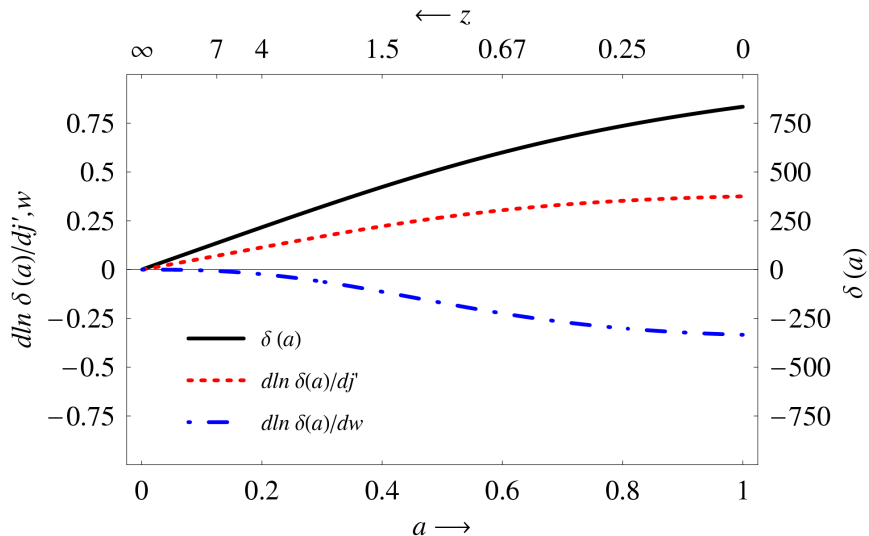}}
\caption{Perturbation $\delta(a)$, and the derivative $d\ln
\delta(a)/dj',dw$, evaluated at $j'=0 (w=-1)$.}
\label{fig:perturbation}
\end{center}
\end{figure}
\section{Discussion}
The above calculations suffice to contrast the $j$ and $w$
formalisms. The procedure that is actually followed in fitting
various data sets is to explore the likelihood of a set of
cosmological parameters subject to prior assumptions about the
nature of the solution. It is clear that sensitivity of a given
analysis to a specific parameter depends upon the source redshift
distribution and such choices can either flatter or disparage a
given approach! The two choices contrasted here are, in the
absence of any compelling theoretical guidance, essentially
arbitrary and it is not our purpose to advocate one over the
other. In particular, it seems to be a mistake to suppose that
because the $w$ formulation produces maximal deviation from
$\Lambda$CDM at $a\sim0.6$ then observations should be
concentrated around $z\sim0.7$. Rather, we seek to raise awareness
of alternative parameterizations of the cosmological model which
may help increase understanding, and prevent misunderstanding, in
the face of forthcoming data.

Of course, just as one should consider performing dark energy
observations over the whole available redshift range, one should also
consider a range of models when analysing these data. Attempts to
measure functions $d(a)$ directly from the data via some necessary
interpolation scheme form one such class of models. The much simpler
jerk and dynamical parameterisations described above trade flexibility
for ease of interpretation, but are as valuable and valid as analytical
tools. The problem of which model provides the most appropriate
description of these data has a well-defined solution that depends on
both the goodness-of-fit (via the likelihood) and the model complexity
(via the prior). The jerk parameterisation is a natural way to perturb
the model that we know to fit the current data very well, and so may be
expected to come with some sensible prior for the parameter $j'$.
Indeed, noting that it is a straight line gradient, insisting on
rotational invariance in the plotting plane leads to a Lorentzian prior
pdf of unit width~\cite{Dose2003}.  This distribution is broad enough
to allow the data to speak for themselves, and featureless for as long
as we ignore any dynamical considerations.  The situation in the
dynamical modelling scenario is somewhat different: there, the onus is
on the theorists to provide prior distributions for
physically-motivated  parameters such as $w$, a task that has not been
undertaken in any great detail as yet.

Our main objective in this work has been to provide an alternative
approach to the problem of characterising the cosmological world
model in the presence of a poorly understood  dark energy
component. We may expect simple parameterisations to be of great
use in interpreting the data arriving from the many current and
future dark energy experiments; this is especially true in the
early stages when the signal-to-noise is low.  The jerk formalism
worked through here has a number of advantages; we hope that it
will be considered by observers as a useful parallel analysis
tool, and as such we have provided an example fitting formula for
the co-moving distance. A more comprehensive jerk toolkit, and a
more detailed discussion of points made earlier in this talk, will
be given elsewhere.
\section{Acknowledgments}
This work was supported in part by NSF grant AST-0206286 and by
the U.S. Department of Energy under contract number
DE-AC03-76SF00515. We would like to thank Nina Hall for drawing
our attention to Eddington's (1935)  view on the cosmological
constant.

\vfill\pagebreak
\end{document}